\documentstyle[12pt]{article}
\def\beq{\begin{equation}}
\def\eeq{\end{equation}}
\def\bea{\begin{eqnarray}}
\def\eea{\end{eqnarray}}
\def\np{{\it Nucl.Phys. }}
\def\pl{{\it Phys.Lett. }}

\def\AP{{\it Ann.Phys.(NY) }}
\def\prl{{\it Phys.Rev.Lett. }}
\def\nc{{\it Nuovo Cimento }}
\def\zp{{\it Z.Physik }}
\def\ijmp{{\it Int.J.Mod.Phys. }}
\def\pr{{\it Phys.Rev. }}
\def\a{\alpha}
\def\t{\theta}
\def\tpr{\theta^{\prime}}
\def\tp{\theta _{+}}
\def\tm{\theta _{-}}
\def\g{\gamma}
\def\ap{\alpha _{+}}
\def\am{\alpha _{-}}
\def\lp{\lambda _{+}}
\def\lm{\lambda _{-}}
\def\lpm{\lambda _{\pm}}
\def\lmp{\lambda _{\mp}}
\def\axx{(\xi _{+},\xi _{-})}
\def\aZZti{(\tilde{Z}_{+},\tilde{Z}_{-})}
\def\aZZ{(Z_{+},Z_{-})}
\def\aZp{(Z_{+})}
\def\aZ{(Z)}
\def\aZm{(Z_{-})}
\def\aZpm{(Z_{\pm })}
\def\tp{\theta _{+}}
\def\tm{\theta _{-}}
\def\axt{(\xi ,\theta )}
\def\tti{\tilde{\theta}}
\def\tpti{\tilde{\theta_{+}}}
\def\tmti{\tilde{\theta_{-}}}
\def\xti{\tilde{\xi}}
\def\D{\Delta}
\def\ipm{_{\pm}}
\def\im{_{-}}
\def\ip{_{+}}
\def\axp{(\xi_{+})}
\def\axpm{(\xi_{\pm })}
\def\Apb{A_{+}^b}
\def\Apf{A_{+}^f}
\def\apb{\alpha_{+}^b}
\def\apf{\alpha_{+}^f}
\def\npF{:e^{\eta F}:}
\def\npFZ{:e^{\eta F(Z)}:}
\def\npFZZ{:e^{\eta (F(Z)+F(Z^{\prime}))}:}
\def\npFZpr{:e^{\eta F(Z^{\prime})}:}
\def\Zpr{Z^{\prime}}
\def\zpr{z^{\prime}}
\def\pp{\partial _{+}}
\def\pzi{(\partial _{z})^{-1}}
\def\ppi{(\partial _{+})^{-1}}
\def\Dp{D_{+}}
\def\Dpi{(D_{+})^{-1}}
\begin {document}
\begin{titlepage}
April 1994 \hfill HU Berlin--IEP--94/7  \\
\mbox{ }  \hfill hep-th@xxx/9404091
\vspace{6ex}
\Large
\begin {center}
\bf{On Super-Liouville Operator Constructions}
\footnote{to
appear in the proceedings of the \em XXVII. Int. Symp. on the Theory of
Elementary Particles \\
\hspace*{6cm} Wendisch-Rietz, September 1993 \em}
\end {center}
\large
\vspace{3ex}
\begin{center}
H.-J. Otto \footnote{e-mail: otto@ifh.de}
\end{center}
\normalsize
\it
\vspace{3ex}
\begin{center}
Math.-nat.Fak. I  der Humboldt--Universit\"at \\
Institut f\"ur Physik \\
Invalidenstra\ss e 110, D-10115 Berlin, Germany
\end{center}
\vspace{6 ex }
\rm
\begin{center}
\bf{Abstract}
\end{center}
\vspace{3ex}
We review the construction of field operators of the N=1 supersymmetric
Liouville theory in terms of the components of a quantized free superfield.
\end {titlepage}
\newpage
\setcounter{page}{1}
\pagestyle{plain}
\section {Classical N=1 super-Liouville theory}

In the following we give a short review on the main features of the N = 1
supersymmetric extension of the previously discussed bosonic Liouville operator
constructions \cite{W}. The derivation of the Liouville action from the
Super-Weyl anomaly of a superstring in d dimensions \cite{Pol} in
"superconformal gauge" of the super-zweibein fields goes quite parallel to the
bosonic model.  Similiar to that case the prefactor $\g ^{-2}$ of the Liouville
action is shifted from its generic value $(10-d)/8\pi $ to $(9-d)/8\pi $ by
taking into account effects from the nontrivial functional measure of the
super-Weyl (or Liouville) field itself \cite{GN1,DK2,DO}.  Here $d$
denotes the embedding dimension of the coupled $N=1$ superstring theory or
$d=\frac{2}{3}c$ where $c$ is the Virasoro central charge of a coupled
superconformal matter theory.  The effective action for
the Liouville superfield $L$ induced this way is (up to a numerical factor)
\beq
S_{L,N=1}=\g ^{-2}\int d^2Z_{+}d^2Z_{-}(D_{+}L D_{-}L -i\mu e^L)~~~~ \big(\g ^2
=\frac{8\pi}{9-d}\big) \label{e1}
\eeq The appearance of a "super-"-cosmological term $\propto \mu $ is not as
uncontroversially motivated as in the bosonic case, as there is no divergence
of
that type in the anomaly and therefore no unescapable need to consider $\mu
\neq
0 $ \cite{Mart} .
We are going to insist on $\mu \neq 0 $
as there is no real argument for not having it, either, and to stay as close as
possible at the bosonic case.

We use the following conventions ($\t _a $ are odd Grassmann variables)
\bea
\xi _a&=&\frac{i}{2}(\tau +a\sigma )~~~~~(a=+,-) \nonumber \\
Z_a&=&(\xi _a,\t _a) \nonumber \\
d^2Z_a&=&d\xi _a d\t _a \nonumber \\
D_a&=&\frac{\partial}{\partial \t _a}+\t _a \frac{\partial}{\partial \xi _a}
\equiv \partial _{\t _a}+\t _a\partial _a
{}~~~~(D_a^2\equiv \partial _a)
\label{e2}
\eea
{}From the action (\ref{e1}) the super-Liouville equation emerges:
\beq
D_{+}D_{-}L\aZZ +i\mu e^{L\aZZ } = 0
\label{e3}
\eeq
Using the component representation for the superfield L
\beq
L\aZZ =\phi \axx +\tp \lp \axx +\tm \lm \axx +\tp \tm a\axx
\label{e4}
\eeq
and eliminating the auxiliary field $a\axx $ the super-Liouville equation

(\ref{e3}) is equivalent to the following coupled system for the bosonic
($\phi $) and fermionic ($\lpm $) components.
\bea
\partial ^2 \phi +\mu ^2e^{2\phi}&=&-i\mu \lp \lm e^{\phi}~~~~~
(\partial ^2 \equiv -\partial _{+}\partial _{-}\equiv \partial _{\tau}^2
-\partial _{\sigma}^2) \nonumber \\
\partial _{\pm}\lmp &=&\pm i\mu \lpm e^{\phi}
\label{e5}
\eea
As we need the most general solution for quantization, we follow the
strategy that proved to be viable in the purely bosonic case
($\lpm \equiv 0$) already:
Choose a simple solution $\phi _0\axx $, transform the variables
analytically (i.e. $\xi _{\pm}\rightarrow A_{\pm}(\xi _{\pm})$
and take into account the (inhomogeneous) transformation
behaviour of the Liouville field under this map. Here we have to generalize
to superanalytic maps both in the "+" and the "-" sectors \cite{Fried} :
\beq
Z=\axt \rightarrow \tilde{Z}=(\xti \axt ,\tti \axt )
\label{e6}
\eeq
with the constraints (sometimes called super Cauchy-Riemann eqs.)
\beq
D\ipm \xti \ipm  =\tti \ipm D\tti \ipm
\label{e7}
\eeq
corresponding to a homogeneous behaviour of the supercovariant derivatives
and differentials
(see (\ref{e2})).
\bea
D\ipm &=&(D\ipm \tti \ipm ) \tilde{D\ipm } \nonumber \\
d^2Z\ipm &=&(D\tti \ipm )^{-1} d^2\tilde{Z\ipm }
\label{e8}
\eea
The transformation law for $L$ ensuring the invariance of (\ref{e1},\ref{e3})
under (\ref{e6}) is given by
\beq
e^{L\aZZ }=(D_{+}\tpti )(D_{-}\tmti ) e^{\tilde{L}\aZZti } .
\label{e9}
\eeq
This characterizes $(\exp L)$ as a superconformal primary field with weights
$(\D _{+}, \D _{-})$=(1/2, 1/2).
A simple solution of the super-Liouvile equation easily is written
down in terms of a "distance" in superspace defined generally as
\beq
Z-Z^{\prime }=\xi -\xi ^{\prime }-\t \t ^{\prime }.
\label{e10}
\eeq
One easily verifies, that one solution of (\ref{e3}) $L_0\aZZ $ is given by
\beq
e^{-L_0\aZZ }=
i\mu (Z\ip -Z\im ).
\label{e11}
\eeq
Two independent superanalytic maps
\beq
Z\ipm =(\xi \ipm ,\t \ipm )
\rightarrow \tilde{Z}\ipm =(A\ipm (Z\ipm ), \a \ipm
(Z\ipm ))
\label{e12}
\eeq
provide a parametrization of the general solution of
(\ref{e3}), where the superfields called $\a $ are Grassmann odd.
Taking eq.(\ref{e9}) into account, one finds \cite{Arv,Bab}
for $L\aZZ =\tilde{L}_0\aZZti $
\beq
e^{-L\aZZ }=i\mu
\frac{A\ip \aZp -A\im \aZm -\a \ip \aZp \a \im \aZm }
{D\ip \a \ip \aZp D\im \a \im \aZm }
\label{e13}
\eeq

with the
superanalyticity constraints (\ref{e7}), i. e.
\beq
D\ipm A\ipm =\a \ipm D\ipm \a \ipm
\label{e14}
\eeq We find it technically more
appropriate to bring (\ref{e13}) by some special "super- M\"obius"
transformation
in the "-"-sector
(essentially $A_{-}\rightarrow -1/A_{-}$) together and some
minor rescalings to the form
\beq
e^{-L\aZZ }=
\frac{1-\mu ^2 A\ip \aZp A\im \aZm -i\mu \a \ip \aZp \a \im \aZm }
{D\ip \a \ip \aZp D\im \a \im \aZm }
\label{e15}
\eeq
This is the complete analog of the "scalar" parametrization used for the
bosonic model \cite{DHJ,Curt,OW} . Again following this analogy, the
unavoidable
choice for a free superfield in terms of which the Liouville field has to be
expressed is
\bea
e^{F\ipm \aZpm }&=&D\ipm \a \ipm  \nonumber \\
D\ip D\im (F\ip +F\im )&=&0
\label{e16}
\eea
For later use we introduce components \beq
F\ipm \aZpm =\psi \ipm \axpm +\t \ipm \chi \axpm
\label{e17}
\eeq
\newpage
\section{Super Liouville Fieldoperators}

Up to now we have written down only classical relations. Going over
to operators defined in the Fock space of the modes of the free superfield
$F\ip +F\im =F$ one tries to find well defined operators of the type

\beq
(e^{-L})^{op}=:e^{-F}:\big[1-\mu ^2A\ip A\im -i\mu \ap \am \big]^{op}
\label{e18}
\eeq
and generally
\beq
(e^{\nu L})^{op}=:e^{\nu F}:\big[(1-\mu ^2A\ip A\im -i\mu \ap \am )^{-\nu }
\big]^{op}
\label{e19}
\eeq
The normal product refers to the free theory ground state.

Similiar to the purely bosonic case we now establish as a construction
principle that the vertex operators (\ref{e19}) are superconformal primaries.
That means, that the corresponding weights
\beq
\D \ip (\nu )=\D \im (\nu )=\frac{\nu }{2}(1-\nu \frac{\g ^2}{4\pi})
\label{e20}
\eeq
derive from the free theory vertex operator $:e^{\nu F}:$ . The free field
theory for $F$ couples to a background charge, leading to an "improvement"
term in the energy-momentum supercurrent and resulting in the term linear in
$\nu $ in
eq. (\ref{e20}) \cite{OTT}. All the operators in brackets $[...]$ in
(\ref{e18},\ref{e19}) must be $(\D \ip =\D \im =0)$-operators. It is the
construction of these "screening operators" in terms of the free field $F$ we
are focusing on in the rest of this talk.

Let us first express the classical superfields $A\ip \aZp ,\a \ip \aZp $ in
terms of $F\ip \aZp $ .
(The corresponding operators in the "$-$" -sector
can be treated in completely the same way.
We are using the well known trick \cite{VEN} of using independent
sets of zero modes $(Q\ip ,P\ip)$ and $(Q\im ,P\im )$
for the "$+$" and the "$-$" sector after replacing $Q\ipm $ by $2Q\ipm $.)

Then at the level of classical fields eq.(\ref{e16}) leads to
\beq
\a \ip =\Dpi (e^{F\ip })=\ppi \big[\Dp e^{F\ip }\big]
\label{e21}
\eeq
We rewrite this in terms of components of $F\ip $ (\ref{e17}) and for
$A\ip ,\a \ip $:
\bea
A\ip &=&\Apb \axp +\tp \Apf \axp    \nonumber \\
\a \ip &=&\apf \axp +\tp \apb \axp .
\label{e22}
\eea
(Here upper indices $b,f$ stand for Grassman even and odd fields,
respectively). We obtain
\bea
\apb \axp &=& e^{\psi \ip \axp }
\label{e23} \\
\apf \axp &=&\ppi \big[e^{\psi \ip }\chi \ip \big] \axp
\label{e24}
\eea
with
\beq
\ppi =const + \int_0^{\xi \ip }d\xi \ip ^{\prime }
\label{e25}
\eeq
and the integration constant has to match the periodicity requirements
both of the integrand and of the integral.

To obtain the bosonic superfield $A\ip \aZp $ we have to integrate the
superanalyticity constraints (\ref{e14}). This results in
\bea
\Apf \axp &=&\apf \axp \apb \axp
\label{e26} \\
\Apb \axp &=&\ppi \big[(\apb \axp )^2+(\pp \apf \axp ) \apf \axp \big]
\label{e27}
\eea
In order to find the operator $(\ap \axp )^{op}$
transforming as a ($\D \ip =0 $)-superfield
the components $\apf ,\apb $ have to be conformal fields with
weights $\D \ip =0$ and $\D \ip =1/2$, respectively. We introduce  therefore a
dressing factor $\eta $ so that (see eqs.(\ref{e1},\ref{e20})
\bea
(\apb \axp )^{op}&=&:e^{\eta \psi \ip \axp }: \\
\label{e28}
\D (\eta )=\frac{1}{2} &=&\frac{\eta }{2}(1-\eta \frac{\g ^2}{4\pi})~~~
\big(\g ^2=\frac{8\pi}{9-d}\big).
\label{e29}
\eea
The resulting solutions for $\eta $ are the conventional ones \cite{Bab,DK2}
for $N=1$
worldsheet supersymmetry :
\beq
\eta \ipm =\frac{1}{4}\big(9-d\pm \sqrt{(9-d)(1-d)}\big)
\label{e30}
\eeq
The operator expressions for
\bea
(\apf )^{op}&=&\ppi \big[(\chi \ip )^{op} (\apb )^{op}\big] \\
\label{e31}
(\Apf )^{op}&=&(\apf )^{op} (\apb )^{op}
\label{e32}
\eea
automatically come up with the correct conformal weights. Short
distance singularities can be rendered integrable (under $\ppi $) by
suitable analytic continuation in
\beq
g=\hbar \eta ^2 \equiv \frac{\g ^2 \eta ^2}{4\pi}=\eta -1
\label{e33}
\eeq
The only more serious problem is posed by the construction of $\Apb \axp $
(27) as conformal primary field with $\D \ip =0$ due to the appearance of
$(\apb \axp )^2$
in the integrand. At operator level this should be translated into
an expression being both well defined and of weight $\D \ip =1$. Both
$[(\apb \axp )^{op}]^2$
(see (\ref{e28})) and $:e^{2\eta \psi \ip \axp }:$
(see (\ref{e29})) are not
acceptable. The brute force way of introducing another "dressing parameter"
$\eta ^{\prime}$ with weight $\D (2\eta ^{\prime})=1$ for the tentative
operator
$:e^{2\eta ^{\prime}\psi \ip \axp}:$  would destroy
supersymmetry and shift the region of direct applicability from $d\le 1$
($\eta $ real) to $d\le -7$ ($\eta ^{\prime }$ real).

A way out is opened up by the fact, that the
leading short distance singularities in the point-splitted version of the
complete integrand of eq.(\ref{e27}) cancel, the remaining short-distance
singularities can be made integrable by analytic continuation in
$g\propto \eta ^2 $ (i.e. in d ) as usual.
As the argument in the component language is too lengthy here,
we give a shorter
but more implicit version in the conformal superfield formalism \cite{Fried}.

Leaving aside the $\pm $ -indices, the problem is to construct a well defined
$\D =1/2$ fermionic superfield operator (i.e. $\D =1/2$ for the lowest
(fermionic)
and $\D =1$ for the bosonic component) for the r.h.s. of
\beq
(DA\aZ )^{op}=(\a \aZ )^{op}(D\a \aZ)^{op}
{}~~~~~( Z\equiv (z,\t ),~~D\equiv \partial _{\t }+\t \partial _z )
\label{e34}
\eeq
on the basis of the bosonic
superfield $(\D =1/2)$ - operator
\beq (D\a \aZ )^{op}=\npFZ
\label{e35}
\eeq
and the fermionic $(\D =0)$ - "screening"-operator
\beq (\a \aZ )^{op}=D^{-1}\npF =\pzi \big[D \npF \big]\aZ
\label{e36}
\eeq
For the time being
we use coordinates on the complex plane
($z=e^{\xi },~~\tau \rightarrow -i\tau $).
Formally, the product of (\ref{e35}) and (\ref{e36}) fits the requirements
for a ($\D =1/2$) - superfield but a closer look immediately reveals the
appearance of the square of $(\apb )^{op}$ (\ref{e28}) being an undefined
operator.  Let us write the inverse of the supercovariant derivative as an
integral operator in superspace:  (The integration constant is irrelevant here)
\beq D^{-1}=\int^z d\zpr \int d\tpr (1+\tpr \t \partial _{\zpr })\equiv
\int^{(Z)}d\Zpr \label{e37}
\eeq
Then we can write formally for the product of (\ref{e35}) and (\ref{e36})
\beq \int^{(Z)}d\Zpr \npFZpr \npFZ
\label{e38}
\eeq
With the propagator
(adapted to the field normalizations used here, continuation
from the euclidean case with radial time ordering is understood)
\beq G(Z,\Zpr )=
-\frac{\g ^2}{4\pi} ln(Z-\Zpr )=
-\frac{\g ^2}{4\pi}\big[ln(z-\zpr )-\frac{\t \tpr }{z-\zpr }\big] \label{e39}
\eeq
we obtain
\beq \npFZpr \npFZ =(Z-\Zpr )^{-g}\npFZZ ~~~~
(g\equiv \frac{(\g \eta )^2}{4\pi}) \label{e40}
\eeq
The bilocal normal product at the r. h. s. can be super-Taylor-
expanded around $Z=\Zpr $ using
\beq
H\aZ =\exp \{ (\Zpr -Z)\partial _y +(\tpr -\t )D_y\} H(Y)_{\vert Y=Z}
\label{e41}
\eeq
Plugging all this into (\ref{e38}), we see that
all the terms with powers ($n\geq 0$, integer) of the
superdistance $(Z-\Zpr )^{n-g}$ integrate to zero, while the surviving
terms are of the form
$$
\int ^zd\zpr (z-\zpr )^{n-g} \times (regular operator)
$$
and can be treated by the usual continuation in g. The short
distance factor corresponds to the well known power of some sinus-
function of the difference in the "cylindrical" $\sigma $-variables (\ref{e2})
\cite{Curt,OW,Nic} due to the map
$z=exp(\xi )\propto exp(\frac{i\sigma}{2})$.

The explicit operator construction of the bosonic and fermionic Liouville
operators in terms of their free field counterparts can now go ahead
parallel to the bosonic construction. What makes things quantitatively
even more complex, is the consideration of different periodicity
conditions for the Neven-Schwarz and Ramond sectors in both chiral
sectors of the free theory, respectively.
At the present (low) stage of
applicability even of the much simpler bosonic Liouville operator formalism a
motivation for working out all the details in the supersymmetric case, too, is
not obvious.

\end{document}